\documentclass[11pt]{article}
\usepackage{amssymb}
\usepackage{amsmath}
\renewcommand{\arraystretch}{1.6}

   \renewcommand{\theequation}{\thesection.\arabic{equation}}
   \csname @addtoreset\endcsname{equation}{section}
    \csname @addtoreset\endcsname{figure}{section}
    \csname @addtoreset\endcsname{table}{section}

\newcounter{thanksnum}
\def\thanksnumber#1
{\setcounter{thanksnum}{\value{footnote}}\setcounter{footnote}{#1}%
                     \addtocounter{footnote}{-1}\footnotemark
                     \setcounter{footnote}{\value{thanksnum}}}
\def\newtheoremz#1{\@ifnextchar[{\@othmz{#1}}{\@nthmz{#1}}}

\def\@nthmz#1#2{%
\@ifnextchar[{\@xnthmz{#1}{#2}}{\@ynthmz{#1}{#2}}}

\def\@xnthmz#1#2[#3]{\expandafter\@ifdefinable\csname #1\endcsname
{\@definecounter{#1}\@addtoreset{#1}{#3}%
\expandafter\xdef\csname the#1\endcsname{\expandafter\noexpand
  \csname the#3\endcsname \@thmcountersepz \@thmcounterz{#1}}%
\global\@namedef{#1}{\@thmz{#1}{#2}}\global\@namedef{end#1}{\@endtheoremz}}}

\def\@ynthmz#1#2{\expandafter\@ifdefinable\csname #1\endcsname
{\@definecounter{#1}%
\expandafter\xdef\csname the#1\endcsname{\@thmcounterz{#1}}%
\global\@namedef{#1}{\@thm{#1}{#2}}\global\@namedef{end#1}{\@endtheoremz}}}

\def\@othmz#1[#2]#3{\expandafter\@ifdefinable\csname #1\endcsname
  {\global\@namedef{the#1}{\@nameuse{the#2}}%
\global\@namedef{#1}{\@thmz{#2}{#3}}%
\global\@namedef{end#1}{\@endtheoremz}}}

\def\@thmz#1#2{\refstepcounter
    {#1}\@ifnextchar[{\@ythmz{#1}{#2}}{\@xthmz{#1}{#2}}}

\def\@xthmz#1#2{\@begintheoremz{#2}{\csname the#1\endcsname}\ignorespaces}
\def\@ythmz#1#2[#3]{\@opargbegintheoremz{#2}{\csname
       the#1\endcsname}{#3}\ignorespaces}

\def\@thmcounterz#1{\noexpand\arabic{#1}}
\def\@thmcountersepz{.}
\def\@begintheoremz#1#2{ \trivlist \item[\hskip \labelsep{\bf #1\ #2}]}
\def\@opargbegintheoremz#1#2#3{ \trivlist
      \item[\hskip \labelsep{\bf #1\ #2\ (#3)}]}
\def\@endtheoremz{\endtrivlist}

\newtheorem{theorem}{Theorem}[section]
\newtheorem{lemma}{Lemma}[section]

\newtheorem{definition}{Definition}[section]
\newtheorem{remark}{Remark}[section]

\newtheorem{example}{Example}[section]

\def\e{\varepsilon}

\def\defi{\stackrel{{\scriptscriptstyle \Delta}}{=}}

\def\a{\alpha}
\def\d{\delta}
\def\o{\omega}
\def\O{\Omega}
\def\q{q}

\def\F{{\cal F}}
\def\w{\widehat}
\def\Ind{{\,\rm Ind\,}}
\def\Ind{{\mathbb{I}}}

\def\R{{\bf R}}
\def\E{{\bf E}}
\def\P{{\bf P}}

\def\b{\beta}
\def\s{\delta}

\def\ww{\widetilde}

\def\t{\theta}
\def\oo{\bar}
\def\s{\sigma}

\def\G{\Gamma}

\def\U{{\cal U}}

\def\M{{\cal M}}

\def\u{u}

\newcommand{\be}{\begin{equation}}
\newcommand{\ee}{\end{equation}}
\newcommand{\bd}{\begin{displaymath}}
\newcommand{\ed}{\end{displaymath}}
\newcommand{\ba}{\begin{array}{ll}}
\newcommand{\ea}{\end{array}}
\newcommand{\baa}{\begin{eqnarray}}
\newcommand{\eaa}{\end{eqnarray}}
\newcommand{\baaa}{\begin{eqnarray*}}
\newcommand{\eaaa}{\end{eqnarray*}}
\font\sm=cmr10

\def\oo{\bar}

\def\Q{{\cal Q}}

\def\q{{\bf q}}
\def\k{{\kappa}}
\def\BB{ A}
\date{Submitted: May 8,  2011. Revised: April 14, 2014}
 \title{
The structure of optimal portfolio strategies  for continuous time
markets  
 }
\author{
Nikolai Dokuchaev
\\ {\sm Department of Mathematics \& Statistics,
Curtin University,}\vspace{-0.5cm} \\\vspace{-0.5cm}  {\sm  GPO Box U1987, Perth, 6845 Western
Australia} \vspace{-0.0cm} \\ {\sm  email N.Dokuchaev@curtin.edu.au}}
\begin{document}
\maketitle
\begin{abstract} The paper studies problem of continuous time optimal portfolio
selection for a incomplete market diffusion model. It is shown
that, under some mild conditions,  near optimal strategies for
investors with different performance criteria can be constructed
using a limited number of fixed processes (mutual funds), for a
market with a larger number of available risky stocks. In other
words, a dimension reduction is achieved via a relaxed version of the Mutual Fund Theorem.
\index{, Mutual Fund Theorem, continuous
time market, near optimal strategies, suboptimal strategies, $\e$-optimal.}
\index{{\bf JEL classification}: C61,
D52, 
D81, 
D84, 
G11 }
\end{abstract}
\section{Introduction}
We study an optimal portfolio selection problem  for a continuous
time stochastic market model which consists of a risk--free bond or
bank account and a finite number of risky stocks.  The evolution of
stock prices is described by Ito stochastic differential equations
with the vector of the appreciation rates $a(t)$ and the volatility
matrix $\s(t)$, while the bond price is exponentially increasing
with a random risk free rate $r(t)$.
\par
These  dynamic  portfolio selection problems are usually studied in
the framework of optimal stochastic control; see, e.g., books of
Krylov (1980) and Fleming and Rishel (1975). There are many works
devoted to different modifications of the portfolio problem (see,
e.g., Merton (1969) and review in   Karatzas and
Shreve (1998)). To suggest a strategy, one needs to forecast future
market scenarios (or the probability distributions, or the future
distributions of $r(t)$, $a(t)$ and $\s(t)$).
 Unfortunately, the
nature of financial markets is such that the choice of a hypothesis
about the future distributions is not easy to justify.
\par
To overcome
limited predictability of the market parameters,
some special methods were developed for the financial models.
One of these tools is the so-called Mutual Fund Theorem which, in the
classical  version, says that the distribution of the risky assets
in the optimal portfolio does not depend on the investor's risk
preferences (or performance criteria). This implies dimension reduction for the optimal portfolio selection
problem:  all
rational investors may achieve optimality using the same mutual fund
plus a saving account. Clearly, calculation of the optimal portfolio
is easier in this case. So far, this property has no analog in
classical stochastic control.
\par
 The Mutual Fund Theorem was established first for the
 discrete time single period mean variance portfolio selection problem, i.e., for the problem with quadratic criteria
(Markowitz (1959)). This
 result was a cornerstone of the modern portfolio theory; in
 particular, the Capital Assets Pricing Model (CAPM) is based on it.
 For the multi-period discrete time setting, some versions of
the Mutual Fund Theorem were obtained so far for problems with quadratic
criteria only (Li and Ng (2000), Dokuchaev (2010)). For the
continuous time setting, the Mutual Fund Theorem was obtained for
portfolio selection problems  for more general utilities. The Mutual
Fund Theorem holds  for  utility functions $U(x)=\d^{-1}x^\d$ and
$U(x)=\log(x)$  for the case of random totally unhedgeable
coefficients, i.e.,  for the case of random coefficients independent on
the driving Brownian motion
 (Karatzas and Shreve (1998)).
It is also known that the Mutual Fund Theorem  does not hold for power
utilities if the coefficients depend on the driving Wiener process (see, e.g., Brennan
(1998)).  Khanna and Kulldorff (1999) proved that the Mutual Fund
Theorem theorem holds for a general utility function $U(x)$ in the
case of non-random coefficients, and for a setting with consumption.
Dokuchaev (2014) extended this result on the case of random totally unhedgeable
coefficients.
Lim and Zhou (2002) found some cases where the Mutual Fund Theorem
theorem holds for problems with quadratic criteria. Dokuchaev and
Haussmann (2001)
 found that the Mutual Fund Theorem holds if the scalar value
$\int_0^T|\t(t)|^2dt$ is non-random, where $\t(t)$ is the market
price of the risk process. In maximin setting, the Mutual Fund Theorem was
established in Dokuchaev (2008,2013).
 Schachermayer {\it et al} (2009) found
sufficient conditions for the Mutual Fund Theorem expressed via
replicability of the European type claims $F(Z(T))$, where
$F(\cdot)$ is a deterministic function and $Z(t)$ is the discounted
wealth generated by the log-optimal optimal discounted wealth
process. The required replicability has to be achieved by trading of
the log-optimal mutual fund with discounted wealth $Z(t)$. It can be summarized that  the Mutual Fund Theorem was established so
far only for several special  optimal portfolio selection problems
and  special market models.
 \par
It appears that there are market models where the classical Mutual Fund Theorem does not hold
but the following relaxed version
of this theorem holds: the optimal  portfolios
with different risk preferences can be constructed using $\mu$
mutual funds only for a market with $n>\mu$ risky stocks.  This $\mu$ can be regarded as a dimension of
the market; in this sense, a market is one dimensional if the
classical Mutual Fund Theorem holds. So far, this feature was studied  for few special settings only. In
particular, single period CAPM models models were studied in a
setting where a number of mutual funds were used to compensate
skewness and consumption (so-called three-moment CAPM, multi-beta
models, or multifactor CAPM); see, e.g., Merton (1973), Poncet
(1983), Fama (1996), Nguyen {\em et al} (2007). A diffusion model
where optimality can be achieved for strategies using  two mutual funds   was discussed in Ingersoll (1987), Chapter
13. In this book,  the optimal strategy was expressed via solution of the Hamilton-Jacobi-Bellman (HJB) equation
(the Bellman equation) for the value function as a  quotient of partial
derivatives of the value function. However, the existence and regularity of these derivatives is difficult to ensure, since the underlying HJB equation is degenerate.
In addition, it is difficult to
ensure that  the resulting stochastic process representing the strategy satisfies reasonable conditions on the
growths such as integrability.  Moreover, it may happen that the quotient found from the HJB equation is not
 smooth enough  to  ensure solvability of the closed loop Ito equations for the
 wealth process.  By these reasons, existence, admissibility, and regularity of the two mutual funds strategy was not yet  established. 
 In theory, this could be overcome  by an alternative martingale approach mentioned briefly in Remark 3.7 in Schachermayer {\em et al} (2009); however, this approach requires replicability of cretain claims and
  does not cover a model with non-hedgeable  Wiener processes.
 
In this paper, we  consider a diffusion market model with
non-hedgeable  Wiener processes and  non-hedgeable  factors such that   the classical Mutual Fund Theorem does not hold. We consider a market with 
$n$ stocks, with $n+N$ independent driving
Wiener processes,  including $N$ non-hedgeable Wiener processes, and with a large number of non-hedgeable factor processes defining the evolution of the market prices.
We found that, for a wide class of utilities, a near optimal (i.e., $\e$-optimal) portfolio
 can be constructed using $\mu<n$
mutual funds only (Theorem \ref{ThM} below). \index{In other words, we  established a {\it
Mutual Funds Theorem}.}  The number $\mu$ is defined by
the number of the non-hedgeable factors correlated with the stock prices, or by
the complexity of correlations in the model, rather than by the
number of stocks or by the total number of random factors. 
\par
The main result (Theorem \ref{ThM}) is obtained under very mild restrictions for the utility functions without
any assumptions on regularity of the value function.
The proof is based on the method of dynamic programming applied indirectly
to some convenient approximations of the original problem
that ensure
certain regularity of the value functions; the range for the strategies is  approximated by bounded sets, and the utility function
is approximated by smooth and bounded functions. This approach has some obstacles: the HJB equations with
bounded admissible controls does not allow explicit solutions. To
overcome these difficulties, we use special time dependent and random constraints
for admissible strategies such that the corresponding HJB admits "almost explicit" solutions generating
near optimal admissible  strategies featuring sufficient regularity and integrability.
\index{An earlier version of this paper was web-published in 2010 [10] and presented in Quantitative Methods in Finance conference in Sydney, Australia, 2011.
The results of this paper were  presented in Quantitative Methods in Finance conference in Sydney, Australia, 2010.}

\section{Model setting}
\label{Sec3DP} We are  given a standard probability space
$(\O,\F,\P)$, where $\O=\{\o\}$ is a set of elementary events, $\F$
is a complete $\s$-algebra of events, and $\P$ is a probability
measure that describes a prior probability distribution.
\par
We assume that the market evolution is driven by a pair of standard
independent Wiener processes
$w(\cdot)=(w_1(\cdot),\ldots,w_n(\cdot))$ and $\w w(\cdot)=(\w
w_1(\cdot),\ldots,\w w_N(\cdot))$ with the values in $R^n$ and $R^N$
respectively. Let $\F_t$ be the filtration generated by $(w(t),\w
w(t))$.
\par
We consider the market model  similar to the model  used in
Dokuchaev (2008, 2013). We assume that the market consists of a risk
free asset or bank account with price $B(t), $ ${t\ge 0}$, and $n$
risky stocks with prices $S_i(t)$, ${t\ge 0}$, $i=1,2,\ldots,n$,
where $n<+\infty$ is given.\par  We assume that \be \label{d.B}
B(t)=B(0)\exp\Bigl(\int_0^t r(s)ds\Bigr), \ee where $r(t)$ is a
$\F_t$-adapted random process of the risk-free interest rate (or the
short rate). We assume that $B(0)=1$. The process $B(t)$ will be
used as numeraire.

The prices of the stocks evolve according to \be \label{d.S}
dS_i(t)=S_{i}(t)\Bigl(a_i(t)dt+\sum_{j=1}^n\s_{ij}(t) dw_j(t)\Bigr),
\quad t>0, \ee where  $a_i(t)$ are the appreciation rates,
$\s_{ij}(t)$ are the volatility coefficients. The initial price
$S_i(0)>0$ is a given non-random constant.
 \par
We  assume  that $r(t)$, $a_i(t)$, and $\s_{ij}(t)$ are  uniformly bounded $\F_t$-adapted measurable random
processes.
\par
We will consider vector processes  $S(t)\defi (S_1(t),\ldots,S_n(t))^\top$  and
$a(t)=(a_1(t),\ldots,a_n(t))^\top$ with the values in
$\R^n$, and a matrix process  $\s(t)\defi\{\s_{ij}(t)\}_{i,j=1}^{n}$
with the values in $\R^{n\times n}$.

Let $\ww S(t)=(\ww S_1(t),\ldots,\ww S_n(t))^\top\defi B(t)^{-1}
S(t)$ be the vector of discounted prices. Let $\ww
a(t)=a(t)-r(t){\bf 1}$, where ${\bf 1}\defi (1,1,...)^\top\in\R^n$.

We assume that the inverse matrix $\s(t)^{-1}$ is defined and
bounded and $r(t)\ge 0$.
\subsubsection*{Wealth and strategies} Let $X_0>0$
be the initial wealth at time $t=0$, and let $X(t)$ be the wealth at
time $t>0$, $X(0)=X_0$.  Let $\ww X(t)\defi B(t)^{-1} X(t)$ be the
discounted wealth.
\par
Let the process $P_0(t)$ be the wealth invested in the
bond, and let $P_i(t)$ be  the wealth invested in the $i$th stock, $i=1,...,n$.
The values of $P_i$ can be negative, in the case of a short position in $i$th
asset.
\par
Let
$\pi_i(t)=B(t)^{-1}P_i(t)$. In this case, the process $\pi_0(t)$
represents the quantity of the bonds, or the discounted wealth
invested in the bond, $\pi_i(t)$, $i\ge 1$, is the discounted wealth
invested in the $i$th stock.
 \par
 We assume that \be \label{XOBS}
\pi_0(t)+\sum_{i=1}^n\pi_i(t)=\ww X(t). \ee
\par
 We denote by $\pi$ the vector process $\pi(t)=\left(\pi_1(t),\ldots
,\pi_{n}(t)\right)^\top$, $t\ge 0$.
\par
\index{\par Let
$$\ww S(t)=(\ww S_1(t),\ldots,\ww S_n(t))^\top\defi
B(t)^{-1}S(t). $$ Let ${\bf S}(t)\defi{\rm diag\,} (
S_{1}(t),\ldots, S_{n}(t))$ and $\ww{\bf S}(t)\defi{\rm diag\,} (\ww
S_{1}(t),\ldots,\ww S_{n}(t))$ be the diagonal matrices with the
corresponding diagonal elements.
\par}
The portfolio is said to be self-financing, if \baaa
\label{in.self1} dX(t)=\sum_{i=1}^n\frac{P_i(t)}{S_i(t)} d
S_i(t)+\frac{P_0(t)}{B(t)}dB(t). \eaaa It can be rewritten as \baaa
\label{in.self11} dX(t)=\sum_{i=1}^n\pi_i(t)^\top \ww{ S}_i(t)^{-1}
d S_i(t)+\pi_0(t)dB(t). \eaaa It follows that for such portfolios
\baa\label{nw}  d\ww X(t)= \sum_{i=1}^n\pi_i(t)\ww S_i(t)^{-1} d\ww
S_i(t)=\pi(t)^\top(\ww a(t)dt+\s(t)dw(t)),\label{wwX} \eaa so $\pi$ alone
suffices to specify the portfolio; see, e.g., Dokuchaev (2007), p. 78.
\par
Let $D$ be the range of the process $\ww X(t)$. We will consider two settings:
with $D=(0,+\infty)$ and with $D=\R$.
\par
We consider a class $\Sigma$  of   admissible strategies consisting
of all $\F_t$-adapted processes
  $\pi(\cdot)=(\pi_1(\cdot),\ldots,\pi_n(\cdot)):[0,T]\times \O\to\R^n$ such
  that the following holds:
  \begin{itemize}
  \item
 If $D=\R$ then  $\sup_{t,\o}|\pi(t,\o)|<+\infty$;
\item
 If $D=(0,+\infty)$ then  $\sup_{t,\o}|\pi(t,\o)|\ww X(t)^{-1}<+\infty$.
\end{itemize}
\par By these definitions, if $D=(0,+\infty)$, then $X(t)>0$ for
any $\pi\in\Sigma$.
\section{The main result}
  Let $T>0$ and $X_0>0$ be
given.
\par
Let $\U$ be the set of all continuous functions $U(\cdot):D\to\R$
such that if $D=\R$ then there exists $c_1>0$  and $c>0$ such that  $|U(x)|\le c_1(1+|x|)^{c}$ for
all $x$.  If $D=(0,+\infty)$, then we
assume that  $|U(x)|\le c_1(|x|^{-c}+|x|^c)$ for some $c_1>0$ and $c>0$.

\index{ there exists $c>0$ such that $\max(0,U(x))\le c(1+|x|)$.
If $D=\R$, then we also assume that $|U(x)|\le c_1(1+|x|)^{c_2}$ for
all $x$, for some $c_1>0$ and $c_2>0$.
 If $D=(0,+\infty)$, then we
assume that  $\min(0,U(x))\ge c_3\log \min(x,1)$ for some $c_3>0$.}

The case where $D=(0,+\infty)$ is included with the purpose to allow
important utility functions with singularity at $x=0$ such as
$U(x)=\ln x$ or $U(x)=-1/x$.

For the sake of generality, we do not exclude
non-differentiable or non-concave $U$. However, discontinuous functions
are not allowed. In particular, step functions used in Dokuchaev and
Zhou (2001)   for the so-called goal achieving problems are not
allowed.  In addition, our setting does not cover utilities with  the exponential growth  such as  $U(x)=-e^{- c x}$ for $D=\R$, $c>0$.

\par
For $U(\cdot)\in \U$, set
 \baaa
V(\pi)\defi\E U(\ww X(T)). \eaaa
\par
We will study the problem \be \label{d.costs0} \mbox{Maximize}\quad
V(\pi) \quad\hbox{over}\quad\pi(\cdot)\in\Sigma.  \ee
\par
Starting from now, we assume that the coefficients $(\ww a,\s)$ are
such that there exist integers $m\ge 0,M\ge 0,N\ge 0$  and
continuous functions \baaa &&{\bf a}:\R^m\times\R^M\times[0,T]\to
\R^n,\qquad {\bf v}:\R^m\times\R^M\times[0,T]\to \R^{n\times n}\eaaa
and functions \baaa {f^\eta}:\R^m\times\R^M\times[0,T]\to
\R^m,\qquad {\b^\eta} :\R^m\times\R^M\times[0,T]\to \R^{m\times
n},\\
{{\w\b}^\eta} :\R^m\times\R^M\times[0,T]\to \R^{m\times N},\\{f^\zeta} :\R^m\times\R^M\times[0,T]\to \R^{M},\qquad
{\w\b^\zeta} :\R^m\times\R^M\times[0,T]\to \R^{M\times N}\eaaa
 such that \baaa
\ww a(t)={\bf a}(\eta(t),\zeta(t),t), \qquad \s(t)={\bf
v}(\eta(t),\zeta(t),t), \eaaa   where  $\eta(t)$ and $\zeta(t)$ are
stochastic processes that take values in $\R^m$ and $\R^M$
respectively and such that they satisfy It\^o equations \baaa
\label{eq1} &&d\eta(t)={f^\eta}(\eta(t),\zeta(t),t) dt +
{\b^\eta}(\eta(t),\zeta(t),t)dw(t)+{{\w\b}^\eta}(\eta(t),\zeta(t),t)
d\w w(t),\\ &&d\zeta(t)={f^\zeta}(\eta(t),\zeta(t),t) dt +
{{\w\b}^\zeta}(\eta(t),\zeta(t),t)d\w w(t). \eaaa Here $\w w(\cdot)$
is a Wiener process with values in $\R^N$ that is independent of $
w(\cdot)$.
\par
The cases where $m=0$, $M=0$, or $N=0$, are not excluded; they represent
models where the corresponding vector processes are absent.
\par
We denote by $|\cdot|$ the Euclidean norm for vectors, the Frobenius
norm for matrices, and the similar norm for elements of the spaces formed as Cartesian products
of spaces of matrices or vectors  such as $\R^n\times \R^{n\times n}$, etc.
\index{$\R^n\times\R^{n\times n}\times\R^m\times\R^{m\times
n}\times\R^{m\times N}\times\R^M\times\R^{M\times N}$.  We denote by
$0_{k\times l}$  the  zero matrix in $\R^{k\times l}$.}
\par We assume
that the following conditions are satisfied:
\begin{itemize}
\item
There exists a constant $C>0$ such that \baaa
&&|F(y_1,z_1,t)-F(y_2,z_2,t)|\le C(|y_1-y_2|+|z_1-z_2|),\\
&&|F(y,z,t)|\le C(1+|y|+|z|) \qquad \forall y_1,y_2,z_1,z_2,y,z,t,
\eaaa where $F=({\bf a},{\bf
v},f^{\eta},\b^{\eta},\w\b^{\eta},f^{\zeta},\w\b^{\zeta})$.
\item We assume that there exists a constant $c_1>0$ such that
$\BB(y,z,t)\BB(y,z,t)^\top\ge c_1I_{m+M}$, where $I_{m+M}$ is the unit
matrix in $\R^{(m+M)\times (m+M)}$, and where the matrix $\BB\in
\R^{(m+M)\times (n+N)}$ is formed as
\renewcommand{\arraystretch}{1.0}\baaa \BB=\left(\begin{array}{cc}
                                               \b^{\eta} & \w\b^{\eta} \\
                                               0_{M\times n} & \w\b^{\zeta}
                                             \end{array}\right).
 \eaaa
 \renewcommand{\arraystretch}{1.6}

 \end{itemize}
\begin{definition} Let $L\ge 1$ be an integer. Consider a set    of $\F_t$-adapted processes
$\M_1(t),...,\M_L(t)$ with the values in $\R^n$.
 Let $\Sigma_{\M_1,...,\M_L}$ be the class of all
 processes $\pi(\cdot)\in \Sigma$ such that
 there exist $\F_t$-adapted one-dimensional processes  $\{\nu_k(t)\}_{k=0}^L$
 such that \baa
 \pi(t)=\sum_{k=1}^L \nu_k(t)\M_k(t).
\label{pinu}
\eaa
\end{definition}
\par
Let $\Q=(\s\s^\top)^{-1}$,  let $\b^\eta_{k}$ be the $k$th column of the matrix $\b^\eta$, and let $\mu=\min(m+1,n)$.
\begin{theorem}\label{ThM}  Consider a
set $\{\M_1(t),...,\M_\mu(t)\}$  of $\F_t$-adapted processes with
values in $\R^n$ defined as \baaa &&\M_k(t)=(\s(t)^\top)^{-1}\b^\eta_{k}(\eta(t),\zeta(t),t),\quad k\le \mu-1, \nonumber\\
&&\M_{\mu}(t)=\Q(t)\ww a(t).\eaaa   For this set, for any
$U(\cdot)\in \U$, \baa \sup_{\pi\in \Sigma} V(\pi)=\sup_{\pi\in
\Sigma_{\M_1,...,\M_\mu}}V(\pi). \label{eqMFT}\eaa
\end{theorem}
\index{\par
Let $\Q(t)=(\s(t)\s(t)^\top)^{-1}$, let
$q_i$ be the $i$th column of the matrix
$(\s(t)^\top)^{-1}=(q_1,...,q_n)$, let $\b^\eta_{ki}$ be the components of the matrix $\b^\eta$, and let $\mu=\min(m+1,n)$.
\begin{theorem}\label{ThM}  Consider a
set $\{\M_1(t),...,\M_\mu(t)\}$  of $\F_t$-adapted processes with
values in $\R^n$ defined as \baaa &&\M_k(t)=\sum_{i=1}^n
q_i(t)\b^\eta_{ki}(\eta(t),\zeta(t),t),\quad k\le \mu-1, \nonumber\\
&&\M_{\mu}(t)=\Q\ww a(t).\eaaa   For this set, for any $U(\cdot)\in \U$, \baa
\sup_{\pi\in \Sigma} V(\pi)=\sup_{\pi\in \Sigma_{\M_1,...,\M_\mu}}V(\pi).\eaa
\end{theorem}}
\vspace{3mm}\par
For the special case of $\mu=1$, $m=0$, $N=0$ (i.e., where the corresponding vector processes are absent), Theorem \ref{ThM} represents   the
relaxed version of the classical Mutual Fund Theorem obtained in
Khanna and Kulldorf (1999) in a setting with consumption and with less general utility functions.
For the case where $\mu=1$, $m=0$, $N>0$, Theorem \ref{ThM} represents  a version of the Mutual Fund Theorem from Dokuchaev (2014).
 A special case where $N=0$ and $M=0$ corresponds to
the  model mentioned in Remark 3.7 in Schachermayer {\em et al} (2009).  A special case where $m=1$ and $M=0$ and where the value function
is regular enough corresponds to
the  model  from Ingersoll (1987), Chapter 13. \par
\section{The implications of Theorem \ref{ThM}}
Let us discuss the  implications and economic interpretation  of Theorem \ref{ThM}.
  Representation (\ref{pinu})  can be interpreted as a distribution of the stock portfolio among $\mu$ mutual funds;  each vector $\M_k(t)$  can be interpreted as a distribution of the stock portfolio  for a mutual fund.  Since the
selection of  $\{\M_k(t)\}$ is independent on  $U(\cdot)$, Theorem \ref{ThM} represents a relaxed version of the Mutual Fund Theorem.
\index{
\begin{remark} {\rm  By the definitions, the supremum of $V(\pi)$ is achieved on the processes $\pi$ represented as
$\pi(t)=\sum_{k=1}^{\mu}\nu_k(t)\M_k(t)$, where  $\nu_k$ are some  $\F_t$-adapted processes with
values in $\R$.  This  can be interpreted as a distribution of the stock portfolio among $\mu$ mutual funds.   Each vector $\M_k$  can be interpreted as a distribution of the stock portfolio, or a stock holding for a mutual fund.
The vector $\M_\mu(t)$ represents the so-called log-optimal portfolio; sometimes, it is called the mean-variance portfolio.
The vectors $\{\M_k(t)\}_{k<\mu}$ represent some hedging portfolios used to compensate correlations in the market.
    The
selection of  $\{\M_k(t)\}$ is independent on
 $U(\cdot)$.  The
selection of the processes  $\nu_k(t)$ depends on
 $U(\cdot)$; these processes are
expressed via solutions of a parabolic Bellman equation in the proof of  Theorem \ref{ThM} below.}
 \end{remark}}
\par
The statement of  Theorem \ref{ThM} can be reformulated as follows:
there exist {\em near optimal} ($\e$-optimal, {suboptimal}) strategies in the class
$\Sigma_{\M_1,...,\M_\mu}$, meaning that, for any $U(\cdot)\in \U$
and any $\e>0$, there exists a strategy $\pi_{U,\e}
\in\Sigma_{\M_1,...,\M_\mu}$ represented as (\ref{pinu}) such that
 \baaa V(\pi_{U,\e})\ge \sup_{\pi\in \Sigma}
V(\pi)-\e. \eaaa
This has a clear economic
interpretation: all investors with different utilities can construct
near optimal strategies by investing in $\mu$ mutual funds only, even
if $n>> \mu$, $M>> \mu$, and $N>> \mu$.
\par
In Theorem \ref{ThM}, the vector $\M_\mu(t)$ represents the so-called log-optimal portfolio; sometimes, it is called the mean-variance portfolio.
For $k<\mu$, the vectors $\M_k(t)$ represent some hedging portfolios used to compensate correlations in the market.
\par
 The processes  $\nu_k(t)=\nu_{U,\e,k}(t)$ for the near optimal strategies   presented in  (\ref{pinu}) depends on
 $U(\cdot)$.   These processes are expressed  in the proof of  Theorem \ref{ThM} below  via derivatives of the smooth approximations of the value functions that are solutions  of some auxiliary  HJB equations.
These equations selected such that their solutions  have the required regularity.
 We emphasize that the statement of Theorem \ref{ThM} itself does not require solvability and regularity of the  HJB equations.
\par
Under very mild conditions on the utility functions,  Theorem \ref{ThM} allows to
reduce  the  original investment problem for a market with $n$ tradable risky
assets to an equivalent  problem for a market with $\mu$ tradable assets. Let us show this.
Consider a matrix process $\M(t)=(\M_1(t), ...,\M_\mu(t))$ with the
values in $\R^{\mu\times n}$ formed from the rows $\M_k(t)^\top$.
Let $\ww a_\xi(t)=\M(t)\ww a(t)$ and $\s_\xi(t)=\M(t)\s(t)$.   Let
us consider a process $\xi(t)=\{\xi_k(t)\}_{k=1}^\mu$ with the
values in $\R^\mu$ defined by the equation \baaa
d\xi(t)={\Xi}(t)(\ww a_\xi(t)dt+\s_\xi(t)dw(t)),\quad \xi_k(0)=1
\quad k=1,...,\mu. \eaaa Here ${\Xi}(t)$ is a diagonal matrix in
$\R^{\mu\times \mu}$ with the diagonal elements ${\Xi}_{kk}(t)=\xi_k(t)$, $k=1,\ldots,\mu$.

Let $\nu(t)=\{\nu_k(t)\}_{k=1}^{\mu}$ be an $\F_t$-adapted process
with the values in $\R^\mu$. Let
$\pi(t)=\sum_{k=1}^{\mu}\nu_k(t)\xi_k(t)=\M(t)^\top \nu(t)$, and let
$\ww X(t)$ be the corresponding discounted wealth.  It follows from
the definitions that \baaa d\ww X(t)=\nu(t)^\top[\ww
a_\xi(t)+\s_\xi(t)dw(t)]= \nu(t)^\top{\Xi}(t)^{-1}d\xi(t). \eaaa
Comparing this with (\ref{wwX}),  we obtain that $\nu(t)$ can be considered as a portfolio
self-financing strategy for a market with the discounted prices
$\{\xi_k(t)\}$ . Therefore, Theorem \ref{ThM} allows to
replace the original investment problem for a market with $n$
stocks by an equivalent  problem for a market with $\mu$ stocks. This could be useful  if
$\mu<<n$.
\begin{remark}{\rm It can be shown that  Theorem \ref{ThM} implies that $|\t(t)|=|\t_\xi(t)|$, where $\t(t)=\s(t)^{-1}\ww a(t)$ is  the market price of risk of the original market, and where $\t_\xi(t)$ is the market price of risk for the reduced market defined as  $\t_\xi(t)=\w\s_\xi(t)^{-1}\ww a_\xi(t)$, where $\w\s_\xi(t)$ 
is a $\mu\times\mu$-dimensional matrix such that  $\w\s_\xi(t)\w\s_\xi(t)^\top=
\M(t)\s(t)\s(t)^\top \M(t)^\top$. Clearly,
if $n=\mu$ then the equality $|\t(t)|=|\t_\xi(t)|$ holds for any non-degenerate matrix $\M(t)$. However, it is interesting to note  that, for $n>\mu$, this equality requires that $\M(t)$ contains
a row proportional to $(\Q^{-1}\w a(t))^\top$ (i.e., such as described in Theorem \ref{ThM}); otherwise, simple counterexamples can be found easily.
This illustrates again  a special role of the log-optimal portfolio $\M_\mu$.}
\end{remark}

 \index{This can be rewritten
as
\baaa
\ww a(t)^\top (\s(t)\s(t)^\top)^{-1}\ww a(t)\equiv \ww a(t)^\top \M(t)^\top (\M(t)\s(t)\s(t)^\top\M(t)^\top)^{-1}\M(t)\ww a(t).
\eaaa}

\index{ It can be noted that if $n=\mu$ and the
matrix $\M(t)$ is invertible  the market price of risk is the same
for the original market model and for the reduced market model. For
the case of a non-degenerate matrix $\M(t)$, it follows from the
equations
 $\s_\xi(t)^{-1}\ww a_\xi(t)=(\M(t)^\top\s(t))^{-1} \M(t)^\top \ww a(t)=\s(t)^{-1}\ww a(t)$.
}
\subsubsection*{Some examples}
It can be noted that our model covers the case where $(\ww
a(t),\s(t))= F(\ww S(t),\eta(t),\zeta(t),t)$, for some deterministic
function  $ F:\R^n\times\R^m\times\R^M\times[0,T]\to \R^n\times
\R^{n\times n}$. It suffices to include the vector $\ww S(t)$ or
some of its components as a part of the vector $\eta(t)$.
\begin{example}
{\rm  Consider a market model where the volatility and the
appreciation rate for  stock prices depend on a market index or
indicator defined by all prices presented in this market. Let $m=1$
and let the market index be $\eta(t)=F(S(t))$, for  some
deterministic function  $ F:\R^n\to \R$, $n>1$; For instance, one
can consider $\eta(t)=\sum_{i=1}^n S_i(t)$. Then $\mu=2$. By
Theorem \ref{ThM}, a  suboptimal strategy can be achieved by
investing in two mutual funds for all risk preferences.
}\end{example}
\begin{example}
{\rm  Consider a market model   such that the volatilities and the
appreciation rates for  stock prices depend on a set of major market
indices such as Dow Jones, FTSE, Hang Seng,  etc. Further, assume
that the movement of the stocks $S_1,...,S_n$ has some impact on one
particular   index, say, on Hang Seng index. For instance, assume
that these stocks are included in this index. This model can be
described as follows: the vector $(\eta(t),\zeta(t))$ represents the
set of market indexes, $m=1$, and the one dimensional process $\eta$
represents the Hang Seng index.  In this case, $\mu=2$. By Theorem
\ref{ThM}, a near optimal strategy can be achieved by investing in
two mutual funds for all risk preferences. }
\end{example}
\begin{example}{\rm
In the previous example,  assume that the dynamics of the stocks
$S_1,...,S_n$ affects $m$ market indexes, say, Dow Jones, Hang Seng,
and some other indexes. In this case, we can use the model with this
$m$ and with $\mu=\min(m+1,n)$. By Theorem \ref{ThM}, a near optimal
strategy can be achieved by investing in $\mu$ mutual funds for all risk preferences.
 }\end{example}

\index{\section*{Appendix: Proofs}\setcounter{equation}{0}
\renewcommand{\theequation}{A.\arabic{equation}}
\renewcommand{\thelemma}{A.\arabic{lemma}}
\renewcommand{\theproposition}{A.\arabic{proposition}}}
\section{Proofs}
\subsection{Reformulation with constrained  strategies}
\begin{definition} Let $K>0$.
 Let $\Sigma(K)$ be the class of all
 strategies $\pi(\cdot)\in \Sigma$ such that
  \begin{itemize}
  \item
 If $D=\R$ then  $\sup_{t,\o} \pi(t,\o)^\top \s(t,\o)\s(t,\o)^\top \pi(t,\o)\le
 K$; and
\item
 If $D=(0,+\infty)$ then  $\sup_{t,\o}\pi(t,\o)^\top \s(t,\o)\s(t,\o)^\top \pi(t,\o)\ww X(t)^{-1}\le K$.
\end{itemize}
In addition, let
$\Sigma_{\M_1,...,\M_L}(K)=\Sigma_{M_1,...,\M_L}\cap \Sigma(K)$, for
a set  $\M_1,...,\M_L$  of $\F_t$-adapted processes with values in $\R^n$.
\end{definition}
\par
Clearly,  $\Sigma=\cup_{K>0}\Sigma(K)$ and
$\Sigma_{\M_1,...,\M_L}=\cup_{K>0}\Sigma_{\M_1,...,\M_L}(K)$.
Therefore, it suffices to prove that
 \baa
\sup_{\pi\in \Sigma(K)} V(\pi)=\sup_{\pi\in
\Sigma_{\M_1,...,\M_\mu}(K)} V(\pi)\quad \forall K>0. \label{eqK}
\eaa
 In this case,
(\ref{eqK}) implies  (\ref{eqMFT}).
\par
Further, Theorem \ref{ThM} holds if  $m+1>n$. In this case, it
suffices to take processes $\M_k(t)=(0,...,0,1,0,...,0)$, with $k$th
component equal to one, $k\le n$. Obviously, any $\pi(t)$ can be
represented
 as a linear combination of these vectors. Therefore, it suffices to
 assume that $\mu=m+1<n$.
\par
 Let us prove (\ref{eqK}). Starting from now, we assume that $K>0$ is given and $\mu=m+1<n$.
\subsection{Some auxiliary lemmas}
 Let  $\Delta(y,z,t)\defi\{u\in\R^n:\
 u^\top {\bf v}(y,z,t){\bf v}(y,z,t)^\top u\le K\}.$
 \par
 Let a  matrix $\w A(u,y,z,t)$ that takes values in $\R^{(1+m+M)\times
(n+N)}$ be defined as
\renewcommand{\arraystretch}{1.0}\baa
\w A(u,y,z,t)=\left(\begin{array}{ccc}\vspace{-0mm} u^\top{\bf v} &
0_{1\times N} \\ \vspace{-0mm}
                                              \b^{\eta} & \w\b^{\eta} \\\vspace{-0mm}
                                              0_{M\times n} & \w\b^{\zeta}
                                              \end{array}\right).
 \label{Am}\eaa\renewcommand{\arraystretch}{1.6}
\begin{lemma}\label{lemmaA}  Let $\G=\{\xi\in \R^{1+m+M}:\ |\xi|=1\}$. For  any $(y,z,t)$,
$$\inf_{\xi\in\G}\sup_{u\in \Delta(y,z,t)} \xi^\top
\w A(u,y,z,t)\w A(u,y,z,t)^\top\xi >0.$$
\end{lemma}
\par
{\it Proof of Lemma \ref{lemmaA}.} It suffices to  replace the supremum over $u$ by the
supremum over $u=\w u$
such that $\w u$ is on the boundary of $\Delta$ and $\b^\eta {\bf
v}^\top\w u=0$. Clearly, this $\w u$ exists since $n>1$ and $m< n-1$.  In
this case,
\renewcommand{\arraystretch}{1.0}
\baaa \w A(\w u,y,z,t)\w A(\w u,y,z,t)^\top= \left(\begin{array}{ccc} \w u^\top{\bf v}{\bf v}^\top \w u & 0_{1\times (m+M)} \\
                                               0_{(m+M)\times 1} & \BB\BB^\top
                                              \end{array}\right)
                                              =\left(\begin{array}{ccc} K & 0_{1\times (m+M)} \\
                                               0_{(m+M)\times 1} & \BB\BB^\top
                                              \end{array}\right). \eaaa
\renewcommand{\arraystretch}{1.6}
By the assumptions on $\BB$, it follows that there exists a constant
$c_1>0$ such that
$$\w A(\w u,y,z,t)\w A(\w u,y,z,t)^\top\ge c_1I_{1+m+M}\qquad \forall y,z,t,$$ where $I_{1+m+M}$ is the
unit matrix in $\R^{(1+m+M)\times (1+m+M)}$. Hence
$$\sup_{u\in \Delta(y,z,t)} \w A(u,y,z,t)\w A(u,y,z,t)^\top\ge c_1I_{1+m+M}\qquad \forall y,z,t.$$
This completes the proof of Lemma \ref{lemmaA}. $\Box$
 \def\bb{{\rm b}}
\begin{lemma}\label{lemmaOpt}
Let $\a\in\R$, $\bb\in\R^n$, $c>0$ be given. Consider the problem:
\baa \hbox{Maximize}\quad -\a|p|^2+p^\top \bb \quad \hbox{over}\quad
p\in\R^n\quad \hbox{subject to}\quad |p|^2\le c. \eaa Then an
optimal solution $p$ exists and the following holds:
\begin{enumerate}
\item If $\a<0, \bb=0$, then any $p$ such that $|p|^2=c$ is optimal.
\item
 If either $\a\ge 0$ or $\a<0, \bb\neq 0$, then the optimal
solution can be selected such that there exists $k=k(\a,\bb,c)\in\R$
such that $p=k\bb$.
\end{enumerate}
\end{lemma}
\par
{\it Proof.} Existence of optimal $p$ follows from the fact that the
domain $\{p:\ |p|\le c\}$ is compact. Statement (i) is obvious. Let
us prove statement (ii). If  $\a=0$ and $\bb\neq 0$ then
$p=\sqrt{c}\bb/|\bb|$ is optimal.

If  $\a=0$ and $\bb=0$ then $p=\bb=0$ is optimal along with all
other admissible $p$.

Let $\a>0$. It suffices to consider the case $\a=1/2$ only.

Clearly, the maximum of the function $g(p)=-|p|^2/2+p^\top \bb$ is
achieved for $p=\bb$. It follows that if $|b|^2\le c$ then $p=\bb$ is
an optimal solution.

If $|\bb|^2>c$ then $p=\sqrt{c}\bb/|\bb|$ is an optimal solution. It can be
seen from the following: \baaa \max_{p: |p|^2\le c} g(p)=
\max_{s\in[0,\sqrt{c}]}\max_{p: |p|=s}g(p). \eaaa Obviously, $ \max_{p:
|p|=s}g(p)=-s^2/2+|\bb|s$  and it is achieved for $p(s)=s\bb/|\bb|$.
The maximum of $-s^2/2+|\bb|s$ over $s\in[0,\sqrt{c}]$ is achieved for
$s=\sqrt{c}$. Hence $p=\sqrt{c}\bb/|\bb|$ is an optimal solution for this case.

Finally, let $\a<0$ and $\bb\neq 0$. Clearly,  $p=\sqrt{c}\bb/|\bb|$ is
optimal again in this case. This completes the proof of the Lemma
\ref{lemmaOpt}. $\Box$ \subsection{Near optimality of constrained Markov strategies}{
 Portfolio
selection problem (\ref{d.costs0}) can be rewritten as \baa
&&\hbox{Maximize}\quad \E U(\ww X(T))
\quad\hbox{over}\quad\pi(\cdot)\in\Sigma\quad\hbox{subject to}
\nonumber\\
&&d\ww X(t)=\pi(t)^\top[{\bf a}(\eta(t),\zeta(t),t)dt + {\bf
v}(\eta(t),\zeta(t),t)dw(t)],\quad
\nonumber\\
&&d\eta(t)={f^\eta}(\eta(t),\zeta(t),t) dt +
{\b^\eta}(\eta(t),\zeta(t),t)dw(t)+{{\w\b}^\eta}(\eta(t),\zeta(t),t)
d\w w(t),\nonumber\\ &&d\zeta(t)={f^\zeta}(\eta(t),\zeta(t),t) dt +
{{\w\b}^\zeta}(\eta(t),\zeta(t),t)d\w w(t), \label{S}\eaa given
$X(0),\eta(0),\zeta(0)$.
\par
It can be seen that, to Markovianize the problem, it suffices to use
the state variables $\ww X(t), \eta(t)$, and $\zeta(t)$.
\par
The following is an adaptation of Definition 3.1.3 from Krylov
(1980), p. 131.
  \begin{definition}
\label{defMar}  Let $\Sigma_M$ be the class of all $\F_t $-adapted
 processes $\pi(\cdot)\in\Sigma$ such that
 there exists a measurable function $u:\R\times \R^m\times
 \R^M\times [0,T]\to \R^n$
 such that
\baaa &\pi(t)=u(\ww X(t),\eta(t),\zeta(t),t)\quad&\hbox{if}\quad
D=\R, \\&\pi(t)=u(\ww X(t),\eta(t),\zeta(t),t)\ww
X(t)\quad&\hbox{if}\quad D=(0,+\infty). \eaaa A process
$\pi(\cdot)\in \Sigma_M$ is said to be a {\em Markov} strategy.
\end{definition}
\begin{remark}{\rm
Note that, by the definition of a Markov strategy, the function
$u(\cdot)$ is such that the closed-loop solution $(\ww
X(t),\eta(t),\zeta(t))$ of Ito equation exists in the class of
$\F_t$-adapted process. Therefore, it may happen that a measurable
and bounded function $u(\cdot)$ does not define   a Markov
strategy.}
\end{remark}
\par
Let $\Sigma_{M}(K)=\Sigma_M\cap \Sigma(K)$. Clearly,
$\Sigma_{M}=\cup_{K>0}\Sigma_{M}(K)$.
\subsection{The proof of Theorem \ref{ThM}}
Note that  the matrix  $A$ defined by (\ref{Am}) represents the
diffusion coefficient for the system of Ito equations in (\ref{S})
for Markov strategies.

Let us first proof the theorem for some special cases.
\subsubsection*{Proof for  bounded  $U$, $U'(x)$, $U''(x)$ and for $D=\R$}
Let us assume that $D=\R$ and the function $U$ is bounded in $D$
together with the derivatives $U'(x)$ and $U''(x)$.
 Set \baa
 J(x,y,z,t)
 \defi
\sup_{\pi(\cdot)\in\Sigma(K)}\E\Bigl\{U(\ww X(T))\Bigl| (\ww X(t),
\eta(t),\zeta(t))=(x,y,z)\Bigr\}. \label{J}\eaa It follows from
Lemma \ref{lemmaA} and Theorem 5.2.5 from Krylov (1980), p.225,
that \baa
 J(x,y,z,t)
 \defi
\sup_{\pi(\cdot)\in\Sigma_M(K)}\E\Bigl\{U(\ww X(T))\Bigl| (\ww X(t),
\eta(t),\zeta(t))=(x,y,z)\Bigr\}. \label{JM}\eaa
\par
 The Bellman equation
 formally satisfied by the value function
 $J=J(x,y,z,t)$ is  \baa
G(t,x,y,z,J'_t,J_{\xi}',J_{\xi\xi}'')=0,\qquad J(x,y,z,T)=U(x),
\label{BelEq}\eaa Here $J_{\xi}'$ the gradient of $J$ with respect
to the vector $\xi=(x,y,z)$, $J_{\xi\xi}''$ is the matrix second
order derivative with respect to the vector $\xi=(x,y,z)$. The
function $G:[0,T]\times \R\times\R^m\times\R^M\times
\R^{1+m+M}\times \R^{(1+m+M)\times (1+m+M)} \to \R$ is defined as
\baaa
G(t,x,y,z,J'_t,J'_{\xi},J''_{\xi\xi})=\sup_{u\in \Delta} G_0(t,u,x,y,z,J'_t,J'_{\xi},J''_{\xi\xi}) + G_1(t,x,y,z,J'_t,J'_{\xi},J''_{\xi\xi}),
\eaaa
where
\baaa
&&G_0(t,u,x,y,z,J_t,J'_{\xi},J''_{\xi\xi})= J_x'
u^\top {\bf a}+{\scriptstyle \frac{1}{2}}  J_{xx}''u^\top {\bf
vv}^\top\u + \, \mbox{tr }[J_{ x y}''u^\top{\bf v}{\b^\eta}^\top]\eaaa
and
\baaa
G_1(t,x,y,z,J'_t,J'_{\xi},J''_{\xi\xi})&=&J_t'+ J_y' {f^\eta}+ J_z'
{f^\zeta}\\& +&  {\scriptstyle
\frac{1}{2}}\mbox{tr }[J_{yy}''({\b^{\eta}\b^{\eta}}^\top +{{{\w\b}^\eta}{{\w\b}^\eta}}^\top)] + \mbox{tr }[ J_{yz}''{{{\w\b}^\eta}
{{\w\b}^\zeta}}^\top]
 +{\scriptstyle \frac{1}{2}}\mbox{tr
}[J_{zz}''{{{\w\b}^\zeta{\w\b}^\zeta}}^\top] .
\eaaa
\par
In this equation, $x\in D$; the set $\Delta$ and the coefficients
depend on $(y,z,t)$. \par Note that $\Delta(y,z,t)$ is a convex set
for all $K,y,z,t$. \par
 By Lemma \ref{lemmaA} and by
Theorem 4.7.4 from Krylov (1980), p. 206, there exists  a unique
solution $J$ that is bounded in any bounded domain together with the
derivatives presented in this equation. By Lemma \ref{lemmaA}
again and  by Theorem 4.7.7 from Krylov (1980), p. 209, it follows
that the function $J$ defined by (\ref{J}) is the solution of
(\ref{BelEq}); in other words, the Verification Theorem holds.  The
Bellman equation does not include generalized derivatives mentioned
in Theorem 4.7.7 from Krylov (1980) because of the existence of
locally bounded derivatives.
\begin{remark}{\rm Technically, Theorems 4.7.4 and 4.7.7
from Krylov (1980) do not cover the case of non-constant $\Delta
=\Delta(y,z,t)$. However, the extension on this  case is
straightforward for our special setting. For instance, one can
consider  the processes $p(t)=({\s(t)^\top})^{-1}\pi(t)$ to be the
strategies instead of $\pi(t)$. In this case, the restriction
$\{\pi(t):\ \pi(t) \in \Delta\}$ is replaced by the restriction
$\{p(t):\ |p(t)|\le K\}$.}
\end{remark}
\par
Let ${\bf v}=({\bf v}_1,...,{\bf v}_n)$, where ${\bf v}_j$ is the
$j$th column of the matrix ${\bf v}$, and let
${\b^\eta}=(\b^\eta_1,...,\beta^\eta_n)$, where $\beta^\eta_j$ is
the $j$th column of the matrix
$\beta_\eta=\{\beta^\eta_{ki}\}_{k,i=1}^{m,n}$.  We have that \baaa
\mbox{tr }[J_{ x y}''u^\top{\bf v}{\b^\eta}^\top]=
\sum_{i=1}^nu^\top {\bf v}_i J_{x y}'' \b^\eta_i =u^\top
\sum_{i=1}^n {\bf v}_i J_{x y}'' \b^\eta_i=u^\top \sum_{i=1}^n {\bf
v}_i\sum_{k=1}^m J_{x y_k}'' \b^\eta_{ki}\nonumber\\=  u^\top
\sum_{k=1}^m
 J_{x y_k}'' \sum_{i=1}^n {\bf v}_i\b^\eta_{ki}.  \label{A1}\eaaa
\par
 It follows that, for a
given $(u,x,y,z,t)$,\baaa
G_0(t,u,x,y,z,J_t,J_{\xi}',J_{\xi\xi}'')= J_x'
u^\top {\bf a}+{\scriptstyle \frac{1}{2}}J_{xx}''u^\top {\bf
vv}^\top\u + u^\top \sum_{k=1}^m
 J_{x y_k}'' \sum_{i=1}^n {\bf v}_i\b^\eta_{ki}.\eaaa
The maximum for $G_0$ in $u$ is achieved for $\w u={{\bf v}^{-1}}^\top p$,
where  $p={\bf v}^\top u$ is a solution of the optimization
problem \baa \hbox{Maximize}\quad \quad -\nu |p|^2 + p^\top
b\quad\hbox{ over} \quad p\in\R^n
 \quad \hbox{subject to}\quad|p|\le K.
 \label{optp}\eaa
 Here $\nu=\nu(x,y,z,t)$ and $b=b(x,y,z,t)$ are defined as
 \baaa
 \nu=-\frac{1}{2}J_{xx}'', \quad b=b(x,y,z,t)= J_x' {\bf v}^{-1} {\bf a}+\sum_{k=1}^m
 J_{x y_k}'' \sum_{i=1}^n {\bf v}^{-1} {\bf v}_i\b^\eta_{ki}.
 \eaaa
\par
By Lemma \ref{lemmaOpt}, problem (\ref{optp}) has an optimal
solution $p(x,y,z,t)=\k(x,y,z,t)b(x,y,z,t)$, where $\k(\cdot):\R
\times \R^m\times\R^M\times[0,T]\to \R$ can be selected to be  a measurable function; its selection depends on $K$.
Hence  the maximum of $G_0$ is achieved for \baa \w u=\w
u(x,y,z,t)=\k {{\bf v}^{-1}}^\top b= \k \Bigl({{\bf v}^{-1}}^\top
J_x' {\bf v}^{-1} {\bf a}+{{\bf v}^{-1}}^\top\sum_{k=1}^m
 J_{x y_k}'' \sum_{i=1}^n {\bf v}^{-1} {\bf v}_i\b^\eta_{ki}\Bigr).
 \label{A2}
 \eaa
 Let  $Q(y,z,t)=({\bf v}(y,z,t){\bf
v}(y,z,t)^\top)^{-1}$.  Equation (\ref{A2}) can be rewritten as
 \baa\w u= \k \Bigl(J_x' Q {\bf a}
+\sum_{k=1}^m
 J_{x y_k}'' \sum_{i=1}^n Q {\bf v}_i\b^\eta_{ki}\Bigr).
 \label{A3}
 \eaa
Further, let $({\bf v}^\top)^{-1}=(\q_1,...,\q_n)$, where $\q_j$ is
the $j$th column of the matrix $({\bf v}^\top)^{-1}$. We have \baaa
Q {\bf v}=({\bf v}{\bf v}^\top)^{-1}{\bf v}=({\bf v}^\top)^{-1}{\bf
v}^{-1}{\bf v}=({\bf v}^\top)^{-1}=(\q_1,...,\q_n).\eaaa Hence $Q
{\bf v}_i=\q_i$, $\sum_{i=1}^n Q {\bf v}_i\b^\eta_{ki}=\sum_{i=1}^n q_i\b^\eta_{ki}=({\bf v}^\top)^{-1}\b^\eta_k$,  and the maximum of $G_0$ is achieved for
 \baa\w u(x,y,z,t)= \sum_{k=1}^{m+1}
 \oo H_{k}(x,y,z,t)\psi_k(y,z,t),
\label{uA5} \eaa
 where
\baaa \psi_k(y,z,t)= ({\bf v}(y,z,t)^\top)^{-1}\b^\eta_{k}(y,z,t),\quad k\le
m, \qquad \psi_{m+1}(y,z,t)= Q(y,z,t) {\bf
a}(y,z,t),\hphantom{}\eaaa and \baa &&\oo H_k(x,y,z,t)= \k(x,y,z,t)
 J_{x y_k}''(x,y,z,t),\quad k\le
m,\quad \nonumber \\ &&\oo H_{m+1}(x,y,z,t)=\k(x,y,z,t)
J_x'(x,y,z,t).
 \label{A33}\eaa
\par
 Assume that the function
$\w u(x,y,z,t)$ is regular enough in $x$  to ensure solvability of
the closed equation (\ref{S}), for instance, it is Lipschitz  in $x$
uniformly in $(y,z,t)$. In this case, the strategy  $\w \pi(t)=\w
u(\ww X(t),\eta(t),\zeta(t),t)$ is optimal and belongs to the class
$\Sigma_M(K)$. Moreover, $
 \pi(t)=\sum_{k=1}^{m+1}\nu_k(t)\M_k(t)$, where \baaa &&\M_k(t)=\psi_k(\eta(t),\zeta(t),t)=
 (\s(t)^\top)^{-1}\b^\eta_{k}(\eta(t),\zeta(t),t),\quad k\le m, \nonumber\\
&&\M_{m+1}(t)=\psi_{m+1}(\eta(t),\zeta(t),t)=\Q\ww a(t).\eaaa  Here
$q_j$ is the $j$th column of the matrix
$(\s(t)^\top)^{-1}=(q_1,...,q_n)$, and \baaa &&\nu_k(t)=\oo H_k(\ww
X(t),\eta(t),\zeta(t),t),\quad k\le m,\qquad \nu_{m+1}(t)=\oo
H_{m+1}(\ww X(t),\eta(t),\zeta(t),t).\hphantom{} \label{A5}\eaaa
\begin{remark}
The
selection of  $\{\M_k\}$ is independent of $K$ and
 $U(\cdot)$.  The
selection of  $\k(x,y,z,t)$  and $\{\oo H_k\}$ depends on $K$ and
 $U(\cdot)$.
 \end{remark}
 \par
 Therefore,  equality (\ref{eqK})
 for the case where $D=\R$ holds for this case of regular enough $\w u$. Moreover, the
 strategy $\w\pi\in \Sigma_{\M_1,...,\M_{m+1}}(K)$ is optimal in $\Sigma(K)$ for this case.
\par
In the general case, it cannot be guaranteed  that the function $\w u(x,y,z,t)$
providing the maximum for $G_0$
 is regular enough in $x$ to ensure solvability of the closed loop
equation (\ref{S}). In this case, we have to approximate $\w u$ by
regular enough functions. We will follow  Chapter 5 from
Krylov (1980), with some simplifications that are possible because
of the following features of our special setting: (a) The maximum for $G_0$ is achieved for $\w u$ that has the
special form (\ref{uA5}); (b) The regularity of $\oo H_k(x,y,z,t)$ in $x$
is sufficient.
\par
 For $R>0$, let ${\rm C}_R= {\rm
S}_R\times [0,T]$, where ${\rm S}_R$ is the origin-centered ball
with the radius $R$ in $\R\times \R^m\times\R^M$.
 We will
  consider  large enough $R\to +\infty$ and small enough $\e\to 0$, $\e>0$.
  \par
  Let  $\ww
H_{k,\e}(x,y,z,t)= \frac{1}{2\e}\int_{-\e}^{\e}\oo H_k(x+q,y,z,t)dq$, and let \baaa &
u_{\e,R}(x,y,z,t)=\sum_{k=1}^{m+1} \ww
H_{k,\e}(x,y,z,t)\psi_k(y,z,t),\quad &(x,y,z,t)\in {\rm C}_R,\\ &
u_{\e,R}(x,y,z,t)=0, \quad &(x,y,z,t)\notin {\rm C}_R.\eaaa It follows
from the definitions that $
u_{\e,R}(x,y,z,t)= \frac{1}{2\e}\int_{-\e}^{\e}\w u(x+q,y,z,t)dq$, for any $(x,y,z)$ from the interior of ${\rm S_R}$ and for small enough $\e$. Hence
\baa
u_{\e,R}(x,y,z,t)\to \w u(x,y,z,t)\quad\hbox{as}\quad \e\to 0\quad \hbox{for a.e. } (x,y,z,t)\in {\rm C}_R.
\label{ue}
\eaa
 Since the set
$\Delta(y,z,t)$ is convex and contains zero vector, we have
that $u_{\e,R}(x,y,z,t)$ takes
the values in $ \Delta(y,z,t)$.
\par Consider the set of closed-loop strategies  $$
\pi_{\e,R}(t)=u_{\e,R}(\ww X_\e(t),\eta(t),\zeta(t),t).$$ Here $\ww X_\e(t)$ is the corresponding discounted wealth.
By the definitions, these strategies belong to
$\Sigma_{\M_1,...,\M_{m+1}}(K)$. Let us show that they are Markov
strategies.
\par
 Let $\tau_{\e,R}$ be the first exit time of the  process
$(\ww X_\e(t),\eta(t),\zeta(t))$ from ${\rm C}_R$. Since the functions
$u_{\e,R}(x,y,z,t)$ are bounded, they take values in $\Delta(y,z,t)$, and, for every $\e>0$, there exists $c>0$ such that \baaa
|u_{\e,R}(x_1,y,z,t)-u_{\e,R}(x_2,y,z,t)|\le c|x_1-x_2|\quad
\forall x_1,x_2,y,z,t, \quad (x_i,y,z,t)\in {\rm C}_R,\quad i=1,2. \eaaa Therefore, the existence of the unique
strong solution of closed equation (\ref{S}) is ensured for
$\pi_{\e,R}(t)=u_{\e,R}(\ww X(t),\eta(t),\zeta(t),t)$ up to the  time $\tau_{\e,R}$. To
prove (\ref{eqK}), it suffices to show that \baa \sup_{\pi\in
\Sigma(K)} V(\pi)=\sup_{\e>0,R>0} V(\pi_{\e,R}).
\label{eqKe} \eaa
\par
Let us prove (\ref{eqKe}).

\par
For a function  $u(x,y,z,t)$, set   \baaa \rho^u(x,y,z,t)=
G_0(t,\w u(x,y,z,t),x,y,z,J_t,J_{\xi}',J_{\xi\xi}'')-G_0(t, u(x,y,z,t),x,y,z,J_t,J'_{\xi},J''_{\xi\xi}).
\eaaa
This equation can be  rewritten as
  \baa \rho^u(x,y,z,t)=J_x'\w u^\top {\bf a}+{\scriptstyle
\frac{1}{2}}J_{xx}''\w u^\top {\bf vv}^\top\w \u + \w u^\top
\sum_{k=1}^m
 J_{x y_k}'' \sum_{i=1}^n {\bf v}_i\b^\eta_{ki}\nonumber\\ -\{J_x' u^\top {\bf
a}+{\scriptstyle \frac{1}{2}}J_{xx}'' u^\top {\bf vv}^\top \u +
u^\top \sum_{k=1}^m
 J_{x y_k}'' \sum_{i=1}^n {\bf v}_i\b^\eta_{ki}\}.
\label{rho}\eaa
Hence
 \baaa &&|\rho^u(x,y,z,t)|\le (|\w u(x,y,z,t)- u(x,y,z,t)|\\&&+\frac{1}{2}|J_{xx}''||\w u(x,y,z,t)^\top {\bf v}{\bf v}^\top
 \w u(x,y,z,t)- u(x,y,z,t)^\top {\bf v}{\bf v}^\top   u(x,y,z,t)|) h_1(x,y,z,t),
 \eaaa
where \baaa
 h_1(x,u,z,t)=|J_x'||{\bf a}|+ \sum_{k=1}^m
 |J_{x y_k}''| \sum_{i=1}^n |{\bf v}_i||\b^\eta_{ki}|.
\eaaa
Applying an obvious inequality $|\w u^\top {\bf v}{\bf v}^\top \w u -u^\top {\bf v}{\bf v}^\top  u|\le
|( \w u-u)^\top {\bf v}{\bf v}^\top(\w u+   u)|$, we obtain that
\baaa &&|\rho^u(x,y,z,t)|\le |\w u(x,y,z,t)- u(x,y,z,t)|h(x,y,z,t),
 \eaaa
 where
 \baaa
 h(x,u,z,t)=h_1(x,y,z,t)+\frac{1}{2}|J_{xx}''||{\bf vv}^\top|(|\w \u|+|u|).\eaaa
As was mentioned already, by Lemma \ref{lemmaA} and by
Theorem 4.7.4 from Krylov (1980), p. 206,  $J$ is bounded in any bounded domain together with the
derivatives presented in this equation. Hence the function  $h(x,u,z,t)$ is bounded on ${\rm C}_R$.
\par
Let $g(t)=\rho^{u_{\e,R}}(\ww X_\e(t),\eta(t),\zeta(t))$.
\par
By It\^o formula, we have that
\baaa
\E\Ind_{\{\tau_{\e,R}>T\}} U(\ww X(\tau_{\e,R}))+\E\Ind_{\{\tau_{\e,R}\le T\}} J(\ww X(\tau_{\e,R}),\eta(\tau_{\e,R}),\zeta(\tau_{\e,R}),\tau_{\e,R}))
\\
=J(X_0,\eta(0),\zeta(0),0)-\E\int_0^{\tau_{\e,R}}g(t)dt.
\eaaa
Hence
 \baaa
J(X_0,\eta(0),\zeta(0),0) =\E\Ind_{\{\tau_{\e,R}>T\}} U(\ww X(\tau_{\e,R}))+r_1+r_2,
\eaaa where
\baaa
r_1=\E\Ind_{\{\tau_{\e,R}\le T\}} J(\ww X(\tau_{\e,R}),\eta(\tau_{\e,R}),\zeta(\tau_{\e,R}),\tau_{\e,R})),\qquad
r_2=-\E\int_0^{\tau_{\e,R}}g(t)dt.
\eaaa
It suffices to show that, for any $\d>0$, there exists $\e$ and $R$ such that
\baa
 J(X_0,\eta(0),\zeta(0),0) \le \E U(\ww X(\tau_{\e,R}))+\d.
\label{JU}\eaa

Let $\d>0$ be given.

Let $\w\tau_R=T\land \inf\{t\ge 0:\ \eta(t)^2+\zeta(t)^2\ge R^2\}$. Clearly, $\tau_{\e,R}\le \w\tau_R$. Since we have assumed that the matrix $\BB\BB^\top>0$  is uniformly non-degenerate, we have
that  $\P(\w\tau_R\le T)\to 0$ as $R\to +\infty$ uniformly in $\e>0$.

 By Corollary 1 from Zakai (1967), it follows that, for any $m>0$,
 \baaa
 \sup_{y,z}(\E |\ww X_\e(\tau_{\e,R})|^m+
 \sup_{\pi\in\Sigma(K)}\E |\ww X(T,\pi)- \ww X_\e(\tau_{\e,R})|^m)<+\infty,\eaaa where
  $\ww X(T,\pi)$ is the discounted terminal wealth for the strategy $\pi$ given that
  \baaa
  \ww X(\tau_{\e,R} ,\pi)= \ww X_\e(\tau_{\e,R}),\quad  \eta(\tau_{\e,R})= y,\quad
    \zeta(\tau_{\e,R})= z.\eaaa  By the assumptions on $U$, it follows that
    $\sup_\e\E  |J(\ww X(\tau_{\e,R}),\eta(\tau_{\e,R}),\zeta(\tau_{\e,R}),\tau_{\e,R}))|^2<+\infty$.
    Hence
    $r_1\to 0$  as $R\to +\infty$ uniformly in $\e>0$. By the assumptions on $U$ again, we obtain that
 $\E\Ind_{\{\tau_{\e,R}\ge T\}} U(\ww X(\tau_{\e,R}))\to 0$ as $R\to +\infty$ uniformly in $\e>0$.   Hence  $\E\Ind_{\{\tau_{\e,R}>T\}} U(\ww X(\tau_{\e,R}))\to \E U(\ww X(\tau_{\e,R}))$ as $R\to +\infty$ uniformly in $\e>0$.

It follows that there exists $R=\w R$ such that \baaa
|r_1|\le \d/3,\qquad\E\Ind_{\{\tau_{\e,R}>T\}} U(\ww X(\tau_{\e,R}))\ge \E U(\ww X(\tau_{\e,R}))-\d/3  \qquad \forall \e>0.
\eaaa

By the  Lebesgue's Dominated Convergence
Theorem, it follows that $r_2\to 0$ as $\e\to 0$ for the given $R=\w R$.
Let $\e=\w\e$ be selected such that $|r_2|\le \d/3$.
It follows that  (\ref{JU}) holds. Hence (\ref{eqKe}) holds.
 This completes the proof of equality (\ref{eqK})
 for the case where $D=\R$ and  where the functions $U$, $U'(x)$,  and $U''(x)$ are bounded in $D$.
\subsubsection{The proof for bounded  $U$, $U'(x)$, $U''(x)$ and for  $D=(0,+\infty)$}
Let us assume that $D=(0,+\infty)$ and the function $U$ is bounded
in $D$ together with the derivatives $U'(x)$ and $U''(x)$. We
consider the change of variables  $q(t)=\ln \ww X(t)$. Using the Ito
formula, we obtain that this change of variables transfers the
corresponding control problem as
 \baa &&\hbox{Maximize}\quad
\E U(e^{q(T)}) \quad\hbox{over}\quad\ww\pi(\cdot)\quad \hbox{subject
to}\nonumber
\\
&&dq(t)=\ww\pi(t)^\top{\bf a}(\eta(t),\zeta(t),t)dt-\frac{1}{2}\ww
\pi(t)^\top {\bf v}(\eta(t),\zeta(t),t){\bf
v}(\eta(t),\zeta(t),t)^\top \ww\pi(t)\nonumber \\&&\hspace{8cm} +
\ww\pi (t){\bf v}(\eta(t),\zeta(t),t)dw(t)],\quad
\nonumber\\
&&d\eta(t)={f^\eta}(\eta(t),\zeta(t),t) dt +
{\b^\eta}(\eta(t),\zeta(t),t)dw(t)+{{\w\b}^\eta}(\eta(t),\zeta(t),t)
d\w w(t),\nonumber\\ &&d\zeta(t)={f^\zeta}(\eta(t),\zeta(t),t) dt +
{{\w\b}^\zeta}(\eta(t),\zeta(t),t)d\w w(t), \label{Sp}\eaa given
$X(0),\eta(0),\zeta(0)$. We consider here maximization over the
strategies $\ww\pi$ from the class $\Sigma(K)$ defined for $D=\R$, i.e., such that
$\sup_{t,\o}|\ww \pi(t,\o)|<+\infty$.
\par
The proof of equality (\ref{eqK}) repeats the proof given above  for
$D=\R$ with few modifications. Instead of (\ref{J}),  we use \baaa
 J(x,y,z,t)
 \defi
\sup_{\pi(\cdot)\in\Sigma_M(K)}\E\Bigl\{U(e^{q(T)})\Bigl| q(t)=x,\,
\eta(t)=y,\,\zeta(t)=z\Bigr\}. \label{Jp}\eaaa Here $x\in\R$ and
$q(t)=\ln(\ww X(t))$; the maximization is over the class $\Sigma(K)$
defined for $D=\R$. The Bellman equation for $J$ is defined
similarly to the Bellman equation for $D=\R^n$, with $G_0$ replaced
by $G_0-{\scriptstyle \frac{1}{2}} J_{x}'u^\top {\bf vv}^\top\u.$
Respectively, $\nu$ in (\ref{optp})
 has to be defined as $\nu=-\frac{1}{2}(-J'_x+J_{xx}'')$. This gives the proof
  of (\ref{eqK}) where $D=(0,+\infty)$  and the functions
$U$, $U'(x)$, and $U''(x)$ are bounded in $D$.
\subsubsection*{Proof for the general case}
  Consider now  the case where either $D=\R$ or $D=(0,+\infty)$  and where the functions $U$, $U'(x)$,
and $U''(x)$ are not necessarily bounded in $D$.
\par
 Let
$\d>0$, $K>0$, and $\oo\pi\in\Sigma(K)$ be given.

\def\UU{\ww U}
For $L>0$, let $\oo U_L(x)=\max(-L,\min(U(x),L))$, and let  $\oo
V_L(\pi)$ be defined similarly to $V(\pi)$ with $U$ replaced by $\oo
U_L$.  Let us select $L>0$  such that $|\oo V_L(\pi)- V(\pi)|\le
\d/5$ for all  $\pi\in\Sigma(K)$; by the assumptions on $\Sigma(K)$,
this $L$ exists. Further,  for $L_1>0$, $\rho>0$,  let a function
$\UU=\UU_{L,L_1,\rho}:D\to\R$ be  such that $|\UU(x)|\le L+1$ for
all $x\in D$, $|\UU(x)-\oo U_L(x)|\le \rho$ if $|x|<L_1$, and such
the derivatives $\UU'(x)$ and $\UU''(x)$ are bounded in $D$. This
function can be obtained via convolution of $\oo U_L$ with a
smoothing averaging kernel, for instance, such as described in
Krylov (1980), Section II.1. Let $\ww V(\pi)$ be defined similarly
to $V(\pi)$ with $U$ replaced by $\UU$. By the assumptions on
$\Sigma(K)$,  there exists $L_1>0$, $\rho>0$  and $\UU(x)$ such that
$|\ww V(\pi)-\oo V_L(\pi)|\le \d/5$  for all  $\pi\in\Sigma(K)$.

By the theorem proved above for the utilities with the properties featured by $\ww U$, there exists  $\w\pi\in \Sigma_{\M_1,...,\M_\mu}(K)$ such that
$\ww V(\w\pi)\ge \ww V(\oo\pi)-\d/5$. In addition, we have that
\baaa V(\w\pi)\ge\oo V_L(\w\pi)-\frac{\d}{5}\ge \ww V(\w\pi)-\frac{2\d}{5}\ge
\ww
V(\oo\pi)-\frac{3\d}{5}
\ge\oo V_L(\oo\pi)-\frac{4\d}{5}\ge V(\oo\pi)-\d.
\eaaa
Since $\oo\pi$ and $\d$ were selected arbitrary,  the proof of (\ref{eqK}) follows  for the general case.

\index{For $L>0$, let $\oo U_L(x)=\max(-L,\min(U(x),L))$, and let $U_L(x)$
be a function such that  $|U_L(x)|\le L+1$ for all $x\in D$,
$|U_L(x)-\oo U(x)|\le \d/3$ for $|x|<L$, and $U'(x)$ and $U''(x)$
are bounded in $D$. Let $\oo V_L(\pi)$ be defined similarly to
$V(\pi)$ with $U$ replaced by $\oo U_L$, and let $V_L(\pi)$ be
defined similarly to $V(\pi)$ with $U$ replaced by $U_L$.
 Clearly, there exists $L>0$ a
such that $\oo V_L(\pi)\ge V(\pi)-\d/3$.  In addition, there exists
$U_L(x)$ such that $V_L(\pi)\ge \oo V_L(\pi)-\d/3$. By the proof
given above, there exists  a set $\{\M_1(t),...,\M_\mu(t)\}$  of
$\F_t$-adapted processes with values in $\R^n$ and $\pi\in
\Sigma_{\M_1,...,\M_\mu}$ such that $V_L(\w\pi)\ge V_L(\pi)-\d/3$.
In addition, we have that  \baaa V(\w\pi)\ge\oo V_L(\w\pi)-\d/3\ge
V_L(\w\pi)-2\d/3\ge V(\pi)-\d.\eaaa Then the proof of (\ref{eqK})
follows for the case where $D=(0,+\infty)$ and for the general case.}
 Finally, the proof of Theorem
\ref{ThM} follows from (\ref{eqK}). $\Box$
 \begin{remark}{\rm  For a typical case, $\k(\ww
X(t),\eta(t),\zeta(t),t)=-J_{x,x}''(\ww
X(t),\eta(t),\zeta(t),t)^{-1}$ if $D=\R$, or  $
\k(q(t),\eta(t),\zeta(t),t)=(J_{x}'(\ww
X(t),\eta(t),\zeta(t),t)-J_{x,x}''(\ww
X(t),\eta(t),\zeta(t),t))^{-1}$ if $D=(0,+\infty)$. It happens when
the strategy $ \pi(t)=\w u(\ww X(t),\eta(t),\zeta(t),t)$ belongs to
the class $\Sigma_M(K)$ and  such that
$\pi(t)^\top\s(t)\s(t)^\top\pi< K$.
 We use the constraints $\pi(t)^\top\s(t)\s(t)^\top\pi(t)\le K$ as  an auxiliary class of
 near optimal (suboptimal) admissible strategies; the final result does not require these constraints.}
\end{remark}
 \begin{remark} {\rm To calculate the processes $\nu_k(t)$, one  have to find $J$ from a HJB equation. Analytical solutions of these equations are rarely feasible; however, numerical
methods for them are well developed; see, e.g.
Barles and Jakobsen (2002) and the review in  Kushner (1990).}  \end{remark}
\section{Conclusion}
The Mutual Fund Theorem defines the distribution of risky assets for
the optimal strategy. If this theorem holds, then the distribution
is the same for all  risk preferences, and the strategy selection
can be reduced to the selection of a one dimensional process of the
total investment in risky assets. This interesting feature is
presented in portfolio theory only and does not have an analog for
the general theory of stochastic optimal control.
  The
efforts in the existing literature are mostly concentrated on the
extension of the list of models where the Mutual Fund Theorem holds. The
current paper suggests a relaxed version of this theorem to cover models
where the classical Mutual Fund Theorem does not hold. We found
conditions that ensure that the optimal strategy can be represented as a linear
combination  of $\mu$ fixed processes (or $\mu$ Mutual Funds), for a
wide class of risk preferences,  for a model with $n>>\mu$ stocks.
The number $\mu$ is defined by the number of correlations in the
model rather than by the number of stocks.
\subsection*{Acknowledgments}  This work  was supported by ARC grant of Australia DP120100928 to the author.\section*{References}
$\hphantom{xx}$  Barles, G,. and Jakobsen, E.R. (2002). On the convergence rate of approximation schemes
for Hamilton-Jacobi-Bellman equations. Mathematical Modelling and Numerical
Analysis, 36(1),  33--54.

Brennan, M.J. (1998). { The role of learning in
dynamic portfolio decisions.} {\em European Finance Review} {\bf 1},
295--306.
\par
Dokuchaev, N., and Haussmann, U.  (2001). Optimal portfolio
selection and compression in an incomplete market.  {\em
Quantitative Finance}, {\bf 1}, 336-345.

{ Dokuchaev, N.G., and Zhou, X.Y. (2001). Optimal investment strategies with
bounded risks, general utilities, and goal achieving. {\it Journal
of Mathematical Economics} {\bf 35}, iss.2,   289-309.}

Dokuchaev N. (2007). {\it Mathematical finance: core theory, problems, and
statistical  algorithms}. Routledge,  New York.

 Dokuchaev, N. (2008).
 Maximin investment problems for discounted and total wealth.
 {\it IMA Journal Management Mathematics} {\bf 19} (1),
63--74.  

Dokuchaev, N. (2010). Mean variance and goal achieving portfolio for
discrete-time market with currently observable source of
correlations.  {\it ESAIM: Control, Optimisation and Calculus of
Variations} {\bf 16}, Number 3, 635--647.

Dokuchaev, N. (2011). Dimension reduction and Mutual Fund Theorem in maximin setting for bond
  market. {\em Discrete and Continuous Dynamical System - Series B} {\bf 16}, No. 4,
  1039--1053.

 Dokuchaev, N. (2014). Mutual Fund Theorem for continuous time markets with random
coefficients.  {\em Theory and Decision}, {\bf 76}, iss. 2, pp. 179--199.

 Fama, E.F. (1996).
Multifactor portfolio efficiency and multifactor asset pricing. {\em
The Journal of Financial and Quantitative Analysis}, Vol. 31, iss.
4, 441--465.

\
Feldman, D. (2007). Incomplete Information Equilibria: Separation
Theorems and Other Myths. {\it Annals of Operations Research} {\bf
151}, 119--149.

Fleming, W.H.,  and  Rishel, R.W. (1975). {\em Deterministic and
Stochastic Optimal Control}, Springer-Verlag, New York.

Ingersoll, J. (1987). Theory of Financial Decision Making, Rowman \&
Littlefield, Totowa, NJ.

Karatzas, I., and Shreve, S.E. (1998).   {\em Methods of
Mathematical Finance}, Springer-Verlag, New York.

Khanna A., and Kulldorff M. (1999). A generalization of the mutual fund
theorem. {\it Finance and Stochastics} {\bf  3}, 167–185 (1999).

Krylov, N.V.  (1980).  {\em Controlled diffusion processes}.
Shpringer, New York.

 Kushner, H. J. (1990). Numerical methods for stochastic control problems in continuous
time. {\em SIAM Journal on Control and Optimization}  28(5), 999--1048.

Li, D.,  and Ng, W.L. (2000). Optimal portfolio selection:
multi-period mean-variance optimization. {\em Mathematical Finance}
{\bf 10} (3),  387-406.

Lim, A. (2004). Quadratic hedging and mean-variance portfolio
selection with random parameters in an incomplete market. {\it
Mathematics of Operations Research} {\bf 29}, iss.1, 132-161.

Lim, A., and Zhou, X.Y. (2002).
 Mean-variance portfolio selection with random
parameters in a complete market. {\it Mathematics of Operations
Research} {\bf 27}, iss. 1, 101-120.

Merton, R.C. (1973). An intertemporal capital asset pricing model.
{\em Econometrica}, 41, 867-887.

Nguyen, D., Mishra, S., Prakash, A., and Ghosh, D. (2007). Liquidity and
asset pricing under the three-moment CAPM paradigm. {\em Journal of
Financial Research} {\bf 30}, iss.  3, 379--398.

Poncet, P. (1983). Optimum consumption and portfolio rules with
money as an asset. {\em Journal of Banking and Finance} {\bf 7}
231-252.

 Schachermayer, W., Sîrbu, M., and Taflin, E. (2009). In
which financial markets do mutual fund theorems hold true? {\it
 Finance and Stochastics} {\bf 13},
49--77.

Zakai, M. (1967). Some moment inequalities for stochastic integrals and for solutions of stochastic differential equations. {\em
Israel Journal of Mathematics} {\bf 5}, Issue 3, pp. 170-176.

\end{document}